\documentclass[aip,preprint]{revtex4-1}
\usepackage{graphicx}
\usepackage{mathrsfs,bm}
\usepackage{colordvi,color,amsbsy,amsmath,bm,amssymb}
\usepackage{hyperref}       % hyperlinks
\begin{document}

%\preprint{AIP/123-QED}

\title{Machine learning-based vorticity evolution and superresolution of homogeneous isotropic turbulence using wavelet projection} %Title of paper



%\author{Tomoki Asaka} 
\author{Tomoki Asaka} 
\email[]{ask05102569@gmail.com}
%\thanks{}
\affiliation{Department of Materials Physics, Graduate School of Engineering, Nagoya University, Furo-cho, Chikusa-ku, Nagoya, 464-8603, Japan}
%\author{Katsunori Yoshimatsu}
\author{Katsunori Yoshimatsu${^*}$}
\email[]{yoshimatsu@nagoya-u.jp}
%\thanks{}
\affiliation{Institute of Materials and Systems for Sustainability, Nagoya University, Furo-cho, Chikusa-ku, Nagoya, 464-8601, Japan}
\author{Kai Schneider}
\email[]{kai.schneider@univ-amu.fr}
%\thanks{}
\affiliation{Institut de Math\'ematiques de Marseille (I2M), Aix Marseille Universit\'e, CNRS, 39 rue F. Joliot-Curie, 13453 Marseille Cedex 13, France}

\date{\today}

\begin{abstract}
A wavelet-based machine learning method is proposed for predicting the time evolution of homogeneous isotropic turbulence where vortex tubes are preserved. 
Three-dimensional convolutional neural networks and long short-term memory are trained with a time series of direct numerical simulation (DNS) data of homogeneous isotropic turbulence at the Taylor microscale Reynolds number 92. 
The predicted results are assessed by using flow visualization of vorticity and statistics, e.g., probability density functions of vorticity and enstrophy spectra.
It is found that the predicted results are in good agreement with DNS results.
The small-scale flow topology considering the second and third invariant of the velocity gradient tensor likewise shows an approximate match.
Furthermore, we apply the pre-trained neural networks to coarse-grained vorticity data using superresolution.
It is shown that the superresolved flow field well agrees with the reference DNS field and thus small-scale information and vortex tubes are well regenerated.
\end{abstract}

%\pacs{}% insert suggested PACS numbers in braces on next line

\maketitle %\maketitle must follow title, authors, abstract and \pacs


% If in two-column mode, this environment will change to single-column format so that long equations can be displayed. 
%\begin{widetext}
%$$\mbox{put long equation here}$$
%\end{widetext}

% Figures should be put into the text as floats. 
% Use the graphics or graphicx packages (distributed with LaTeX2e).
% Put the label that you will use with \ref{} command in the braces of the \label{} command.
%
% \begin{figure}
% \end{figure}

% \begin{table}
% \end{tabular}
% \end{table}
%---------------------------------------------------------------------------
\section{Introduction}
Self-organization in small-scale turbulence is ubiquitous for instance in the form of coherent vortex tubes.
The vortices defined as regions of intense vorticity magnitude are characteristic at scales in the dissipation range (e.g., Refs. \onlinecite{Jime,KanedaJoT}). 
They are intermittently distributed in physical space.
Moreover, they play key roles for the dynamics of e.g., inertial particle clustering,\cite{BC} mixing in combustion,\cite{Tanahashi} and extreme acceleration of fluid particles.\cite{Porta}
The representative length scale and time scale are respectively the Kolmogorov length scale $\eta$ and the Kolmogorov time scale $\tau_\eta$.
\citet{Jime} found that the typical diameter of the tubes is about $10\eta$.
In turbulence modeling, such as large-eddy simulation, 
the influence of the vortices is statistically modeled without resolving the scales in the dissipation rage.

Machine learning in fluid dynamics is an active rapidly evolving and promising field.
For reviews we refer to Refs. \onlinecite{DuraARFM,brunton2020machine,SchuJoT}. 
The increasing power and capabilities of machine learning approaches can provide benefit to in particular computational fluid dynamics.
Computationally expensive direct numerical simulation (DNS) computations may thus be reduced or even avoided in the near future by training neural networks with available turbulent flow data. \cite{kochkov2021machine,vinuesa2022enhancing}
The application of machine learning covers, e.g., extraction of flow features, turbulence modeling, superresolution (SR) of unresolved flows, and time-evolution of flows. 
Among various applications, we focus in the following on SR and predicting the time evolution of flows.

SR is not limited to improve the resolution of images.\cite{Park2003}
\citet{Dong} proposed a deep-learning SR method, by using convolutional neural networks (CNN) 
which learn the local area of flow structure via %a 
convolution filters. 
SR using two-dimensional (2D) CNN has been applied to turbulence;
2D freely-decaying homogeneous turbulence,\cite{Fukami2D} urban turbulence,\cite{Ohnishi} three-dimensional (3D) forced homogeneous isotropic turbulence (e.g., Refs. \onlinecite{Liu,Kim2020}), and 3D turbulent channel flows (e.g., Refs. \onlinecite{Liu,Kim2020}).
\citet{Liu} trained velocity fields and showed that the SR using a time series of the coarse-grained data as an input is able to reconstruct turbulent statistics better than the SR using a single time input data. 
Generative adversarial networks (GAN) were introduced by \citet{Ledig}
The learning by GAN progresses such that the probability density function (PDF) of the generated data is well superimposed on that of the correct data, and therefore GAN can well reconstruct the statistics of the correct data. 
SR based on an unsupervised learning model using a cycle-consistent GAN (CycleGAN)\cite{Zhu2017} has been proposed in \citet{Kim2020} and applied in the context of homogeneous isotropic turbulence and turbulent channel flow.  
They trained 2D slices of velocity fields and showed that SR using the CycleGAN well preserves statistics of velocity and vorticity.
\citet{SRGue} use a 2D GAN for SR of turbulent velocity fields in channel flows, and proposed a downsampling factor normalized by the wall-unit quantities in the estimate of SR of channel turbulent flows.
Yousif {\it et al.}\cite{GAN2Dto3D} 
%\citet{GAN2Dto3D} 
proposed a GAN-based model for reconstructing 3D turbulent velocity fields from velocity data in 2D planes.
Asaka {\it et al.}\cite{asaka2022wavelet} developed an SR method using wavelets and 3D CNN. 
They showed that SR of the coarse-grained vorticity field reproduces the vortex tubes much better than that of the coarse-grained velocity for homogeneous isotropic turbulence.
Applying wavelets to the flow fields yields a sparse multiresolution representation and thus well reduces the number of degrees of freedom of turbulence at all scales, in particular at the scales in the dissipation range.
The wavelet transform well catches the information of position and scale of the fields and thus reflects spatial locality and neighboring relationships.

For a review of machine-learning-based SR reconstruction for vortical flows, we refer to Ref. \onlinecite{Fukamireview}.
Recently, Refs. \onlinecite{FukamiSSR,GueSSR,SantosSSR}
have developed SR reconstruction from data observed at sparsely distributed positions in physical space
using computational and experimental fluid dynamic data. 
Yousif {\it et al.}\cite{YouTransSR} combined a transformer\cite{trans} with an SR GAN\cite{YSRGAN} for predicting velocity fields of a spatially developing turbulent boundary layer.

As expected the quality of SR becomes worse, 
when the input data get coarser.
We however anticipate that training of time evolution of the vortex tubes can be a key of improvement for SR of the vortex tubes. 
Using DNS of homogeneous isotropic turbulence,
Yoshida {\it et al}.\cite{Yoshida} showed that the time-evolution of the DNS data of small-scale eddies at scales  $k\eta \lesssim 0.2$ are perfectly regenerated from the DNS data of larger-scale eddies at scales $k\eta \gtrsim 0.2$ after some transient time, 
if the latter are assimilated at each time step.
Here, $k$ is the modulus of the wavenumber.

\citet{Hasegawa1, Hasegawa2020} developed a method for predicting %time 
flow time evolution, using reduced order modeling together with 2D CNN, long short-term memory (LSTM) and autoencoder for 2D flows past a cylinder.
LSTM is a neural network which learns stored information over extended time intervals by recurrent back propagation.\cite{LSTM97}
In autoencoders, the encoder reduces the data size, while the decoder reconstructs the data.\cite{AutoE}
The method was extended to 3D turbulent channel flows.\cite{Nakamura2020}
The use of convolutional autoencoders, 3D CNN with LSTM was proposed by %\citet{MohanJoT} 
Mohan {\it et al}.\cite{MohanJoT} 
for different turbulent flows including 3D homogeneous isotropic turbulence. 
\citet{Shanker2022} developed a method for predicting time evolution of 3D homogeneous isotropic turbulent flows,
utilizing continuous neural ordinary differential equations, which was proposed by \citet{NODE} instead of LSTM.
\citet{Guastoni2021} developed proper orthogonal decomposition (POD) based machine learning for 3D turbulent channel flow and well predicted the coefficients of POD modes near the wall region.
Lucor {\it et al}.\cite{lucor2022simple} proposed physics-informed neural networks for modeling turbulent convection. 
These studies train representative quantities at the scales in the energy containing range, 
such as velocity and temperature fluctuation.
The degree of freedoms in small scales are much reduced in the learning.
Mohan {\it et al}.\cite{Mohan0} proposed a deep-learning model of large and inertial scale dynamics in turbulence, by using wavelet thresholding instead of the autoencoder.
\citet{Peng2023} introduced a linear attention based model, which is coupled with Fourier neural operators,\cite{FNO} for 3D homogeneous isotropic turbulence.
The model is trained with vorticity fields at a relatively low Taylor micro-scale Reynolds number of $30$, 
and showed reasonable agreement between their predicted results and the correct data for several large-eddy turnover times.
This method is further developed for turbulence modeling in~\citet{LiArx}

In this paper, we propose a machine learning method using orthogonal wavelets to predict time evolution of vorticity fields in 3D homogeneous isotropic turbulence, which is one of the most canonical turbulent flows, while preserving vortex tubes. 
To this end we combine 3D CNN and LSTM techniques in a concise way to design our machine learning approach, and we train coarse-grained vorticity fields. 
We here use linear wavelet projection, not wavelet thresholding, such that we can track the time-evolution of turbulence in wavelet space without the use of an adaptive wavelet basis for data compression. 
Orthogonal wavelets are suitable for representing turbulent flow fields, which are multi-scale and intermittent. 
Coherent vortex tubes can be thus efficiently retained in wavelet space.
\cite{Farge2001,okamoto2007coherent} 
The training and test data are obtained by DNS of homogeneous isotropic turbulence.
The wavenumber where its enstrophy spectrum hits the maximum is sufficiently larger than the maximum wavenumber in the large-scale external forcing range. 
We assess the proposed wavelet-based method of flow prediction by using, e.g., visualization, PDFs of vorticity, and enstrophy spectra. 
Furthermore, we apply the pre-trained machine learning model to SR in order to predict small-scale vorticity from a coarse-grained vorticity at a given time instant, and assess the results.

The remainder of the manuscript is organized as follows. 
In Sec.~\ref{sec:dns}, we describe DNS of homogeneous isotropic turbulence.
We summarize orthogonal wavelet representation in Sec. \ref{sec:wavelet}.
In Sec. ~\ref{sec:method}, we show a machine learning method for time-evolution of vorticity in homogeneous isotropic turbulence.
The method is based on wavelet projection the vorticity.
The prediction by the machine learning is then assessed in Sec.~\ref{sec:nume_result}. 
An application of the pre-trained model to SR is given in Sec. \ref{SR}. 
Finally, conclusions are drawn in Sec. ~\ref{sec:conclusions}.

%+++++++++++++++++++++++++++++++++++++++
\section{Direct Numerical Simulation}
\label{sec:dns}
%+++++++++++++++++++++++++++++++++++++++
For our machine learning, we use the data of 3D incompressible homogeneous isotropic turbulence obtained by DNS in a periodic domain $\Omega = [0, 2 \pi]^3$.
The turbulent dynamics is governed by the Navier--Stokes equations 
\begin{eqnarray}
\frac{\partial {\bm u}}{\partial t} + ({\bm u} {\bm \cdot} \nabla) {\bm u} = - \frac{1}{\rho} \nabla p + \nu \nabla^2 {\bm u} + {\bm f}, \label{NS}
\end{eqnarray}
where the velocity field is divergence free, $ \nabla {\bm \cdot} {\bm u} = 0$.
Here, ${\bm u}({\bm x},t)$ denotes velocity, $p({\bm x},t)$ pressure,
${\bm f}({\bm x},t)$ the external forcing,
$\rho$ the constant density, $\nu$ the constant kinematic viscosity,
$t$ time, ${\bm x}=(x,y,z)$,
and $\nabla=(\partial/\partial x,\partial/\partial y,\partial/\partial z)$.
To simplify notation, the arguments ${\bm x}$ and $t$ are omitted when useful.
The spatial mean velocity of ${\bm u}$ is set to zero.

DNS has been carried out,
using a Fourier spectral method and a fourth-order Runge--Kutta scheme for time integration. 
Aliasing errors are removed by a phase shift method and a spherical cutoff filter.
Only Fourier modes satisfying $k < \sqrt{2} n_g /3$ are retained,
where $k = |{\bm k}|$,
${\bm k}$ is the wave vector, 
and $n_g$ denotes the number of grid points in each Cartesian direction.
The kinematic viscosity is set to $\nu=2.3 \times 10^{-3}$.
The initial field of the DNS at $t=0$ is a divergence-free random velocity field whose energy spectrum $\propto k^2 \exp(-k^2/4)$, and the spatial average of energy per unit mass, denoted by ${\mathcal{E}}$, is set to $0.5$:
${\mathcal{E}}= (1/2) \langle |{\bm u}|^2 \rangle$, 
where $\langle \cdot \rangle = (2\pi)^{-3} \int_{\Omega} \cdot \, d {\bf x} $.
For the forcing ${\bm f}$, negative viscosity \citep{Jime} was used only for $k<2.5$ such that the spatial average of energy per unit mass, denoted by ${\mathcal{E}}$, remains constant and equals $0.5$. 
The number of grid points $n_g^3$ is $128^3$,
and the time increment is $2.0 \times 10^{-3}$.
The data size and the resulting Reynolds number of the DNS are limited due to the memory requirement imposed by our machine learning model (see Sec. \ref{sec:method}).

\begin{figure}
\begin{center}
\includegraphics[width=0.6\textwidth]{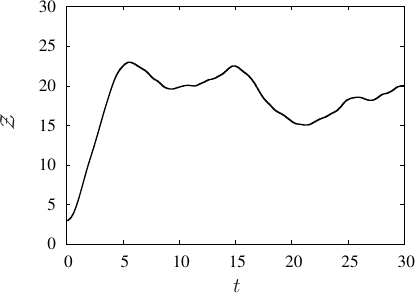}
\end{center}
\caption{Time development of the enstrophy ${\mathcal{Z}}$ of the DNS. }
\label{DNSeps} 
\end{figure}

Figure \ref{DNSeps} shows the time development of the enstrophy ${\mathcal{Z}}$,
defined as ${\mathcal{Z}}=(1/2) \langle |{\bm \omega}|^2 \rangle$,
where ${\bm \omega}$ is vorticity.
We can see that ${\mathcal{Z}}$ becomes statistically quasi-stationary for $t \gtrsim 5$.
Time-averaged statistics of the DNS are summarized in Table~\ref{tab1}.
The average is obtained by using $250$ snapshots at every $0.08$ in the interval $10.00 \le t \le 29.92$.
The time averaged energy dissipation rate per unit density, denoted by ${\bar \epsilon}$, is  the time average of $2 \nu \mathcal{Z}$.
The integral length scale $L$ is defined as $L=\pi/(2 u'^2) \int_0^{k_{\mathrm{max}}} (1/k) {\bar E}(k) dk$, and $u'= \sqrt{2 {\mathcal E}/3}$. 
Here, ${\bar E}(k)$ is the time-averaged energy spectrum 
in which the energy spectrum $E(k)$ at each time instant is obtained by $E(k)=(1/2) \sum'_k \left| {\hat {\bm u}} ({\bm q}) \right|^2$,
where ${\hat {\bm u}}({\bm k})$ is the Fourier transform of ${\bm u}({\bm x})$, 
and $\sum'_k$ denotes the summation over the spherical shell, $k-1/2 \le |{\bm q}| < k+1/2$.
The Kolmogorov microscale $\eta$ is defined as $\eta =(\nu^{3}/{\bar \epsilon})^{1/4}$.
Thus, $k_{\mathrm{max}} \eta \approx 1.16$, where $k_{\mathrm{max}}$ is the maximum wavenumber retained by the DNS. 
The Taylor microscale Reynolds number $R_\lambda$ is defined by $u' \lambda/\nu$, 
where $\lambda = (15 \nu u'^2/ {\bar \epsilon})^{1/2}$.

\begin{table}
\caption{Time-averaged statistics of the $250$ snapshots of the DNS data at $t=10 +0.08 n \, (n=0,\cdots,249)$.
The interval $0.08$ is about $0.5 \tau_\eta$ and $ 0.17 T $.
Here, $T$ is a large-eddy turnover time defined by $L/u'$, 
and $\tau_\eta $ is the Kolmogorov time scale defined by $(\nu/{\bar \epsilon})^{1/2}$.}
\begin{center}
\begin{tabular}{cccccc} \hline
 ${\bar \epsilon}$ & $L$ & $\eta$ & $T$ & $\tau_\eta$ & $R_\lambda$ \\ \hline
$8.57 \times 10^{-2}$ & 1.21 & $1.94 \times 10^{-2}$& 0.476 & 0.164 & 91.9 \\ \hline
\end{tabular}
\end{center}
\label{tab1}
\end{table}

%-------------------------------
\section {Wavelet Representation}\label{sec:wavelet}
%-------------------------------
We introduce the wavelet decomposition of a 3D vector field and represent each component, $ v({\bm x}) \in \ L^2({\Omega})$,  as an orthogonal wavelet series,
where $\Omega=[0,2\pi]^3$.
The field is sampled on $2^J$ grid points in each Cartesian direction, 
where $J$ corresponds to the number of octaves in each space direction in the domain (e.g., $J=7$ for $128$ grid points).
The wavelet functions $\psi_{\mu,j,{\bm i}} ({\bm x})$ yield an orthogonal basis at scale $2^{-j}$, 
where $\mu (= 1, \cdots, 7)$ denotes directions and ${\bm i}= (i_x,i_y,i_z)$ position. 
The scaling function at scale $2^{-j}$ is denoted by $\phi_{j,{\bm i}} ({\bm x})$.
The fast wavelet transform (FWT) is used to compute the wavelet coefficients from the field
and the inverse fast wavelet transform (IFWT) to reconstruct the field from the wavelet coefficients, 
which in some cases, e.g. for coarse-graining, are filtered. 
The computational complexity of FWT and IFWT is $O(2^{3J})$.
Similar to previous work by Refs. \onlinecite{Farge2001,okamoto2007coherent}, 
we use Coiflet 12 wavelets which have filter length 12, four vanishing moments and compact support. 
We use here PyWavelets~\citep{lee2019pywavelets} in which the Coiflet 12 wavelets are denoted by `coif2'.

The field $v({\bm x})$ can be then decomposed into an orthogonal wavelet series applying either a periodization or a folding technique,\cite{mallat1999wavelet} 
the latter in the case of subcubes:
\begin{eqnarray}
v_j({\bm  x}) &=& v_{j-1}({\bm  x}) + w_{j-1}({\bm  x}),   
\end{eqnarray}
where
\begin{eqnarray}
v_j({\bm  x}) &=& \sum_{i_1,i_2,i_3 =0}^{2^{j-1} -1} \langle v, \phi_{j,{\bm i}}  \rangle  \phi_{j,{\bm i}} ({\bm x}), 
\end{eqnarray}
and 
\begin{eqnarray}
w_j({\bm  x}) &=& \sum_{\mu=1}^7 \sum_{i_1,i_2,i_3 =0}^{2^{j-1} -1} \langle v, \psi_{\mu,j,{\bm i}}  \rangle  \psi_{\mu, j,{\bm i}} ({\bm x}),
\end{eqnarray}
where $\left<\cdot,\cdot\right>$ denotes the $L^2$-inner product, defined as $\left< g_1, g_2 \right> = (2\pi)^{-3} \int_{\Omega} g_1({\bm x}) \, g_2({\bm x}) d {\bm x}$. 
Note that $v_J= v $, $v_0=\langle v \rangle$ and $j=1,\cdots, J$.
As the wavelets have vanishing moments, including their mean value, we have $\langle w_j \rangle =0 $.
The wavelet and scaling coefficients are given respectively by $\langle v, \psi_{\mu,j,{\bm i}} \rangle$, and $\langle v, \phi_{j,{\bm i}}  \rangle$.
At scale $2^{-j}$ we have $2^{3j}$ scaling and $7 \times 2^{3j}$ wavelet coefficients.
We recall that the flow fields satisfy periodic boundary conditions.

%+++++++++++++++++++++++++++++++++++++++
\section {Machine Learning Approach}\label{sec:method}
%+++++++++++++++++++++++++++++++++++++++
We develop a methodology for predicting time-evolution of vorticity of 3D homogeneous isotropic turbulence, 
combining 3D CNN, LSTM, and orthogonal wavelet analysis, denoted by WCNN-LSTM here. 
In the following, we describe the training data and its preprocessing in Sec.~\ref{subsec:training}, 
and then present the procedure of WCNN-LSTM in Sec.~\ref{subsec:ml}. 
In Sec.~\ref{subsec:test} we describe the test data and the output data.

%+++++++++++++++++++++++++++++++++++++++
\subsection{Training Data}\label{subsec:training}
%+++++++++++++++++++++++++++++++++++++++
We use $205$ snapshots of the DNS data at every time interval $\Delta \tau$ in $10.00 \le t \le 26.32$ in our machine learning,
where $\Delta \tau = 40 \Delta t$.  
The time step of the DNS is $\Delta t = 2.0 \times 10^{-3}$, 
and here $\Delta \tau $ is about half of the Kolmogorov time $\tau_{\eta}$, i.e., $ \Delta \tau \approx 0.5 \tau_{\eta}$.
The spatial resolution of the DNS data is $128^3$ grid points, as mentioned in Sec. \ref{sec:dns}. 
The time correlation of vorticity is defined as 
$ \langle {\bm \omega}({\bm x},t) {\bm \cdot} {\bm \omega}({\bm x},t+\Delta \tau) \rangle/[\langle |{\bm \omega}({\bm x},t)|^2 \rangle \langle |{\bm \omega}({\bm x},t+\Delta \tau)|^2 \rangle]^{1/2} \approx 0.81$
and the correlation is obtained by the time-average of the snapshots.
Sequential six vorticity fields at every time interval $\Delta \tau$ at $t =  t_0^{(i)} + n \Delta \tau \, (n = -4, -3, \cdots,0,1)$ are randomly chosen.
Here, our batch size is 4, and $t_0^{(i)}$ $(i=1,2,3,4)$ is an initial time of each set labelled by $^{(i)}$. 
The five fields for $n \le 0$ are used as input data,
while one field for $n=1$ , i.e. at $t = t_0^{(i)} + \Delta \tau$, is used as its correct data. 
We here use wavelet projection for compression of the DNS data.
Each snapshot is decomposed into a 3D orthogonal wavelet series applying the FWT with periodic boundary conditions,
decomposing only one level. 
We then obtain wavelet coefficients in seven directions and the scaling coefficients at the coarser scale $2^{-J+1}$. 
Here we have $J=7$.
The sampled values of $v({\bm x})$ are well approximated by the scaling coefficients weighted by $2^{3J/2}$ at scale $2^{-J}$.\cite{mallat1999wavelet}
Our machine learning model trains the weighted coefficients.
In the case of Haar wavelets the projection corresponds to the application of a box filter, 
while for Coiflet 12 we use here the projection corresponds to a more sophisticated low pass filter.

%+++++++++++++++++++++++++++++++++++++++
\subsection{Machine Learning Model}\label{subsec:ml}
%+++++++++++++++++++++++++++++++++++++++
\begin{figure}
\begin{center}
\includegraphics[width=\linewidth,keepaspectratio]{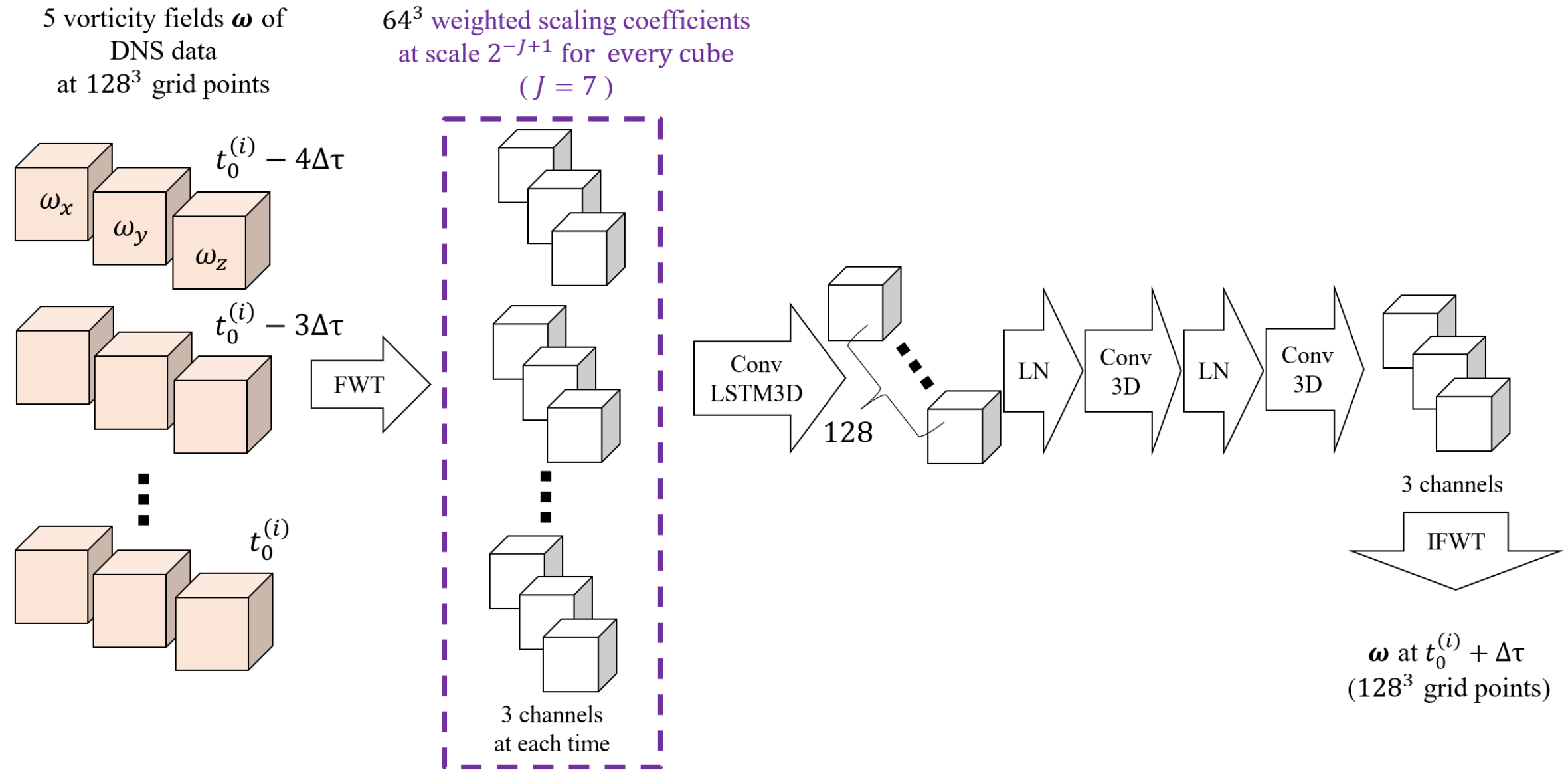}
\end{center}
\caption{Illustration of the structure of the machine learning model, WCNN-LSTM.} 
\label{model}
\end{figure}
\begin{figure}
\begin{center}
\includegraphics[width=7.5cm,keepaspectratio]{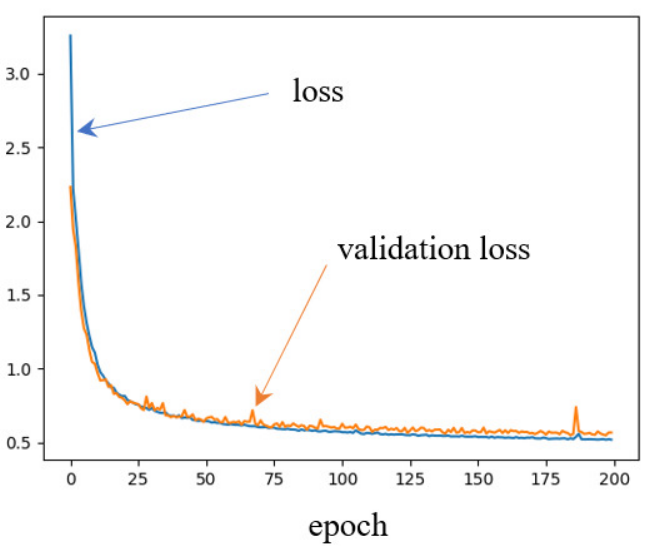}
\end{center}
\caption{Epoch-dependence of the loss and validation loss in WCNN-LSTM.} 
\label{error}
\end{figure}

%-------------------
We illustrate the procedure of our machine learning model WCNN-LSTM in Fig.~\ref{model}. 
The training scheme is implemented using the TensorFlow open-source library\cite{tensorflow2015-whitepaper} together with Python 3.6.8 interaction interface. 
We use ConvLSTM3D layer, LayerNormalization (LN), and Conv3D layer (keras.io) for 3D CNN and LSTM. 
Our machine learning model has seven layers; one input layer, five middle layers and one output layer. 
The first layer is the input layer, while the seventh layer is the output layer.
In the middle layers, we use filters which are 3D kernels with their sizes being $3 \times 3 \times 3$ in order to catch local features of the turbulence data. 
The second layer is ConvLSTM3D layer having $128$ filters. 
The third and fifth layers are LN layers, i.e., for layer normalization. 
The fourth and sixth layers are Conv3D layers having $48$ and $3$ filters, respectively.
Zero-padding is employed at each layer, to keep the size of the output data being 
the same as those of the input data. 
As the activation functions for Conv3D layer, we use the linear function. 
For LSTM we use $\mathrm{tanh}$ as activation function and sigmoid as the recurrent activation function. 
The adaptive moment estimation optimizer, called Adam optimizer\cite{kingma2014adam} is used. 
The number of epochs is set to 200. 
The error is measured by the use of a loss function, 
where the mean-absolute of the difference between the output data and the correct data. 
Figure \ref{error} shows a plot of the loss and the validation loss as a function of the epochs.
The optimal parameters in the model are determined by using $\mathrm{modelcheck}$ such that the validation loss is the smallest in epoch considered here. 
Our learning follows this procedure $15$ times, and then we get 15 possible models.
Each model predicts its output data,
using remaining training data which are not used in the learning of the 15 possible models.
In the following we select the model where the absolute value of the mean of the output data is the lowest.
Note that for the DNS data the spatial average, i.e., the mean of vorticity ${\bm \omega}$ vanishes, $\langle {\bm \omega} \rangle={\bm 0}$. 
The output of TensorFlow and the code of this machine learning is open access and can be found on GitHub \url{https://github.com/KYoshimatsu/WCNNLSTM.git}.

%-------------------
%-------------------

%+++++++++++++++++++++++++++++++++++++++
\subsection{Test Data and Output Data}\label{subsec:test}
%+++++++++++++++++++++++++++++++++++++++
The flow prediction procedure is detailed in the following.
First, we use five vorticity fields obtained by the DNS at time instants,
$t = {\hat t} + n \Delta \tau$ $(n = -4,-3,-2,-1,0)$ as input data,
and predict the field at $t =  {\hat t} + n \Delta \tau$ with $n$ being positive. 
Our test data are five DNS snapshots in the interval $29.60 \le t \le 29.92$ for each $\Delta \tau$. 
Here, ${\hat t}$ is a starting time, and we set ${\hat t}$ to $29.92$.

Our model predicts the scaling coefficients at scale $2^{-J+1}$ ($J=7$) and at time $t = {\hat t} + \Delta \tau$ . 
We then apply IFWT to the coefficients to get the output vorticity data at $t = {\hat t} + \Delta \tau (=30.0)$ in physical space,
while the wavelet coefficients at scale $2^{-J+1}$ are set to zero.
The output data in physical space satisfy periodic boundary conditions.
The input data in the next step are four vorticity fields of DNS at $t =  {\hat t} + n \Delta \tau \, (n = -3, \cdots, 0)$ 
and the field predicted by WCNN-LSTM at $t =  {\hat t} + \Delta \tau $.
Then we predict the vorticity field at $t = {\hat t} + 2\Delta \tau$, using the same procedure as what we have used in getting the output data at $t = {\hat t} + \Delta \tau$. 
We apply this procedure step by step.
The input data in the $\ell$-th step with $\ell \in \mathbb{N}$ are
the data at $t = {\hat t} + n \Delta \tau$ $(n = -5 +
, \cdots,\ell)$.
The data for positive $n$ are the predicted data obtained by WCNN-LSTM at the previous steps,
while the data for $n\le 0$ are the test data.
We took another set of DNS data with different ${\hat t}$ and confirmed that we got results which are qualitatively similar to those shown in Sec.~\ref{sec:nume_result}.

%+++++++++++++++++++++++++++++++++++++++
\section{Machine Learning Results of Vorticity Evolution}\label{sec:nume_result}
%+++++++++++++++++++++++++++++++++++++++
%*************************
\begin{figure}
\begin{center}
\includegraphics[width=12cm,keepaspectratio]{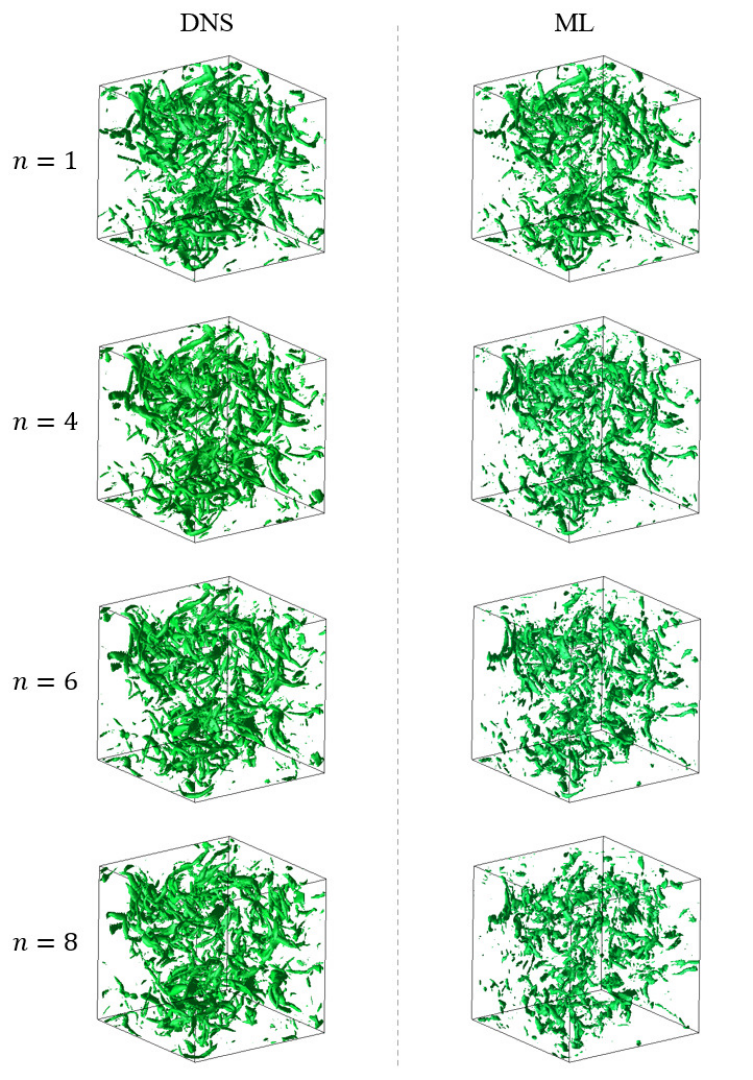}
\end{center}
\caption{Visualization of the isosurfaces of $|{\bm \omega}|$ at $|{\bm \omega}| = \omega_m + 2.5 \omega_\sigma$ at each time instant.
Here $\omega_m$ and $\omega_\sigma$ are the mean value and the standard deviation of the corresponding vorticity modulus $|{\bm \omega}|$ for the DNS data, respectively:
(left) DNS data and (right) ML data, which are our predicted data, at $t =  {\hat t} + n\tau \, (n = 1, 4, 6 $ and $ 8)$. }
\label{vis3D}
\end{figure}
%------------------
\begin{figure}
\begin{center}
\includegraphics[width=13cm,keepaspectratio]{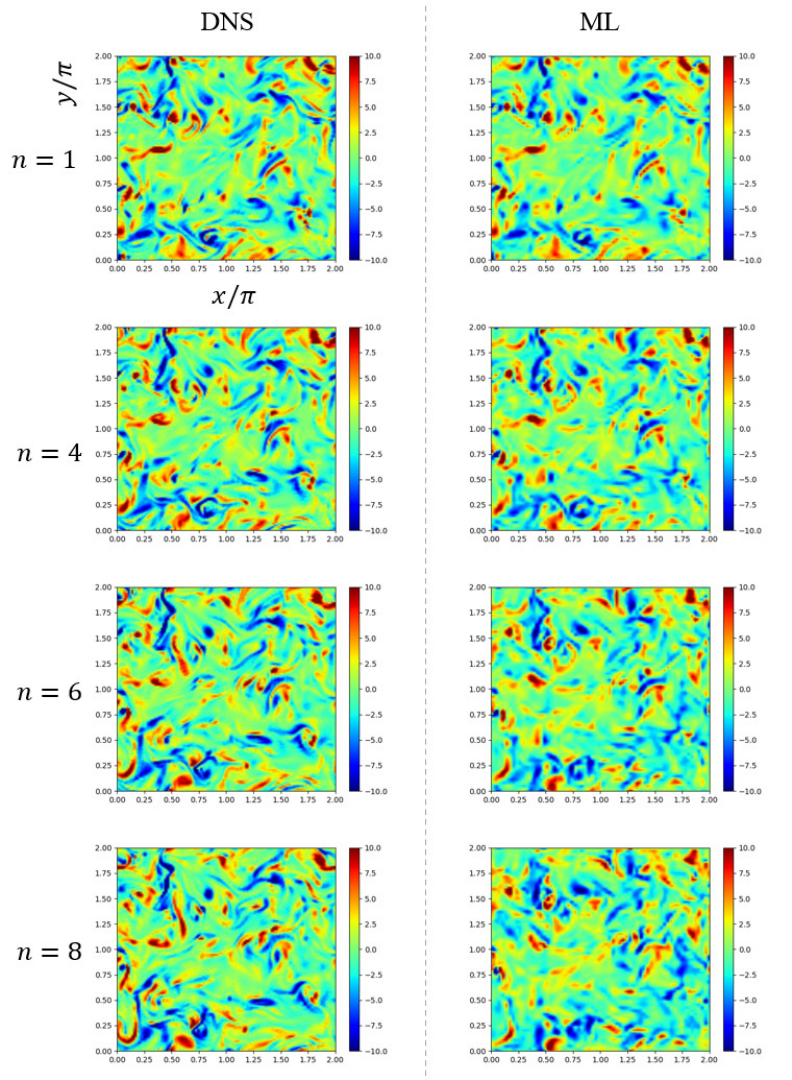}
\end{center}
\caption{Visualization of the vorticity components $\omega_x$ on an $x-y$ plane: (left) DNS data and (right) ML data at $t =  {\hat t} + n\tau \, (n = 1, 4, 6 $ and $8)$.}
\label{vis2D}
\end{figure}
%*************************
In this section, we verify and estimate our WCNN-LSTM flow prediction, using visualization of vorticity and turbulence statistics.
Figure~\ref{vis3D} presents the visualization of the modulus of vorticity $|{\bm \omega}|$ for the DNS data and the predicted data by WCNN-LSTM, at $t =  {\hat t} + n \Delta \tau \, (n = 1, 4, 6 $ and $ 8)$. 
Here we denote the predicted data for simplicity by ML.
Shown are isosurfaces satisfying $|{\bm \omega}| = \omega_m  + 2.5 \omega_\sigma $ with $128^3$ grid points, 
where $\omega_m $ and $\omega_\sigma $ are respectively the mean and standard deviation of $|{\bm \omega}|$ of the DNS data at each time instant. 
We find that pronounced vortex tubes are well preserved by ML,
though we can see some discrepancy between the tubes obtained by DNS and by ML.
The tubes predicted by ML are less intense compared to those obtained by DNS.
To look at the vorticity obtained by the DNS and ML in detail,
we visualize one vorticity component, $\omega_x$, in an $x-y$ plane at different time instants in Fig. \ref{vis2D}.
We can see that the ML vorticity excellently agrees with the DNS vorticity at $n=1$. 
However, the discrepancy between the vorticity fields obtained by DNS and by ML seems to become larger with increasing time, as is expected.
It is to be noted that owing to flow sensitivity of turbulence,
small discrepancy between two statistically identical turbulent flows at a given time instant grows in time.

%*************************
\begin{table}
\caption{Time evolution of enstrophy ${\mathcal Z}$ for DNS data, CG1 data, and ML data at time ${\hat t} + n \Delta \tau$ ($n=1,4,6,8,$ and $10$).} 
\begin{center}
\begin{tabular}{cccccc} \hline
$\mathcal{Z}$ & $t=t_1$ & $t=t_4$ & $t=t_{6}$ & $t=t_{8}$ & $t=t_{10}$\\ \hline
DNS & 20.0 & 20.0 & 20.0 & 19.9 & 19.8 \\ \hline
CG1  & 18.3 & 18.4 & 18.3 & 18.2 & 18.1 \\ \hline
ML  & 17.9 & 17.9 & 17.5 & 17.2 & 17.1 \\ \hline
\end{tabular}
\end{center}
\label{tabZ}
\end{table}
%------------------
\begin{table}
\caption{Time evolution of energy ${\mathcal E}$ for ML data.  
The energy of the DNS is kept constant at $0.5$, as mentioned in Sec. \ref{sec:dns}.
In the CG1 data, ${\mathcal{E}} \approx 0.5$ and therefore we omit them for brevity from Table \ref{tabE}. } 
\begin{center}
\begin{tabular}{cccccc} \hline
$\mathcal{E}$  & $t=t_1$ & $t=t_4$ & $t=t_{6}$ & $t=t_{8}$ & $t=t_{10}$ \\ \hline
ML  & 0.48 & 0.45 & 0.44 & 0.45 & 0.46 \\ \hline
\end{tabular}
\end{center}
\label{tabE}
\end{table}
%*************************

Table \ref{tabZ} presents the enstrophy ${\mathcal{Z}}$ at five time instants for the ML data in comparison with ${\mathcal{Z}}$ for the DNS data and for the coarse-grained data at scale $2^{-J+1}$.
The latter is hereafter denoted by CG1.
We can see that the enstrophy ${\mathcal{Z}}$ for the ML data well agrees with  ${\mathcal{Z}}$ for the DNS at each time instant within the difference ranging from about $10\%$ to $14\%$.
The difference increases with time ${\hat t} + n \Delta \tau$.
The reason is mainly attributed to the wavelet projection at scale $2^{-J+1}$ necessary for getting CG1 data.
We recall that before applying the IFWT the wavelet coefficients at scale $2^{-J+1}$ are set to zero in WCNN-LSTM.
The enstrophy ${\mathcal{Z}}$ of the ML data agrees with ${\mathcal{Z}}$ of the CG1 data much better than ${\mathcal{Z}}$ of DNS.
Table \ref{tabE} gives the energy ${\mathcal{E}}$ for the ML data.
We can see that ${\mathcal{E}}$ of the DNS is well reproduced by the ML at each time.
The difference of ${\mathcal{E}}$ between the ML data and DNS data ranges between $4\%$ to $12\%$. 
The velocity fields of the data are obtained from the predicted vorticity ${\bm \omega}$ by using the Biot--Savart law: ${\bm u} = -\nabla^{-2} (\nabla \times {\bm \omega}) $. 

%*************************
\begin{figure}
\begin{center}
\includegraphics[width=7.5cm,keepaspectratio]{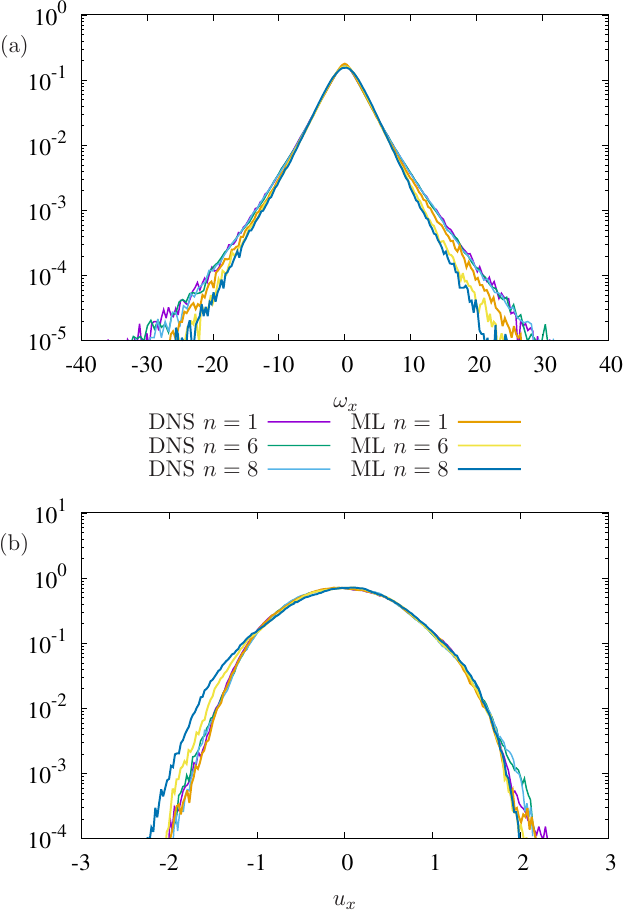}
\end{center}
\caption{PDFs of (a) the $x$-component of vorticity $\omega_x$, and (b) the $x$-component of velocity $u_x$.
We here omit the PDF of $\omega_x$ for CG1 data at $n=1$, because the PDF well agrees with the counterpart of the DNS data. } 
\label{fig:pdf}
\end{figure}
%*************************

We show the PDFs of the $x$-components of vorticity $\omega_x$ and velocity $u_x$ at different time instants in Fig. \ref{fig:pdf}.
In Fig.~\ref{fig:pdf}(a), 
we can see that the PDFs of $\omega_x$ for ML well overlap with those of $\omega_x$ for DNS at each time instant.
We observe some departure between the PDFs for ML and DNS, especially in the stretched tails of the PDFs.
The departure becomes larger as time progresses.
The PDFs for ML are narrower than those for DNS, which implies that the vorticity predicted by ML is less intermittent than that of DNS.
In Fig.~\ref{fig:pdf}(b), 
we can see that at $n=1$, the PDF of $u_x$ for ML well overlaps with that for DNS.
The PDFs are close to a Gaussian distribution.
At $n=6$ and $8$, we see that the PDFs for ML depart from those of DNS.
We omit the PDFs of the $y$ and $z$-components of vorticity and velocity, 
because the observations are the same as in the case of the PDFs of the $x$-components due to statistical isotropy.

%*************************
\begin{figure}
\begin{center}
\includegraphics[width=6cm,keepaspectratio]{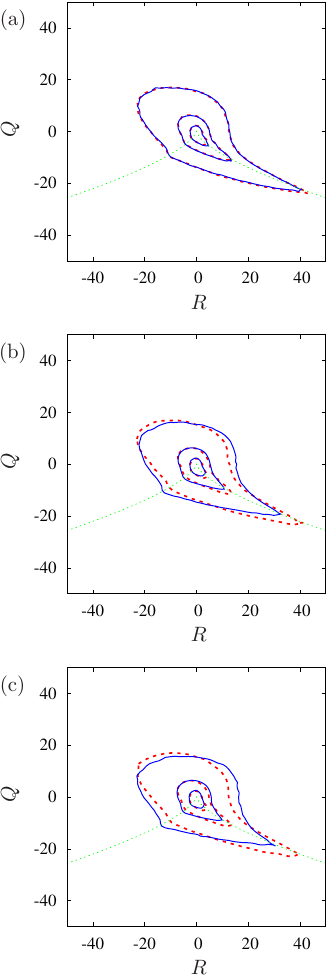}
\end{center}
\caption{Joint PDFs of $Q$ and $R$ for DNS and ML at (a) $n=1$, (b) $n=6$, and (c) $n=8$.
The contour lines for DNS and ML are denoted by the red dashed lines and the blue solid lines, respectively.
The contour lines for all cases are set to $10^{-4}$, $10^{-3}$ and $5 \times 10^{-3}$, starting near the origin. 
The green dotted lines represent $27R^2/4 +Q^3=0$.
} 
\label{jpdf}
\end{figure}
%*************************

In order to get deeper information of the ML data, 
we examine small-scale flow topology.
The topology can be characterized by the second and third invariant of the velocity gradient tensor.\cite{PQR,Soria}
These invariants are respectively defined as $Q = - (1/2) A_{ij} A_{ji}$ and $R=- \mathrm{det} (A_{ij}) $, 
where $A_{ij}=\partial u_i/\partial x_j$ and the Einstein summation convention is used for the repeated subscripts.
Here, we use the notation $(x_1,x_2,x_3)=(x,y,z)$.
It is to be noted that $A_{ii} =0$ for DNS, owing to the divergence-free condition. 
In our ML, $A_{ii} =0$ likewise vanishes, 
because we have used the Biot--Savart law for computing the velocity fields for ML. 
Figure~\ref{jpdf} shows isolines of the joint PDFs of $Q$ and $R$.
The PDFs for DNS are tear-drop like, as reported in e.g., Refs. \onlinecite{Soria,Chevillard}.
Figure \ref{jpdf} (a) shows that the joint PDF of $Q$ and $R$ well agrees with that of DNS at $n=1$.
We can see in Figs.~\ref{jpdf} (b) and \ref{jpdf} (c) for $n=6,8$ that
the isolines of the highest value $5 \times 10^{-3}$ in the joint PDFs for ML are in fairly good agreement with those for DNS, 
though we can also see some discrepancy between ML and DNS PDFs for the other smaller isoline values.
We omit the PDFs of $Q$ and $R$ for CG1 for brevity, because the PDFs for CG1 excellently agree with those for DNS.

%*************************
\begin{figure}
\begin{center}
\includegraphics[width=7.5cm,keepaspectratio]{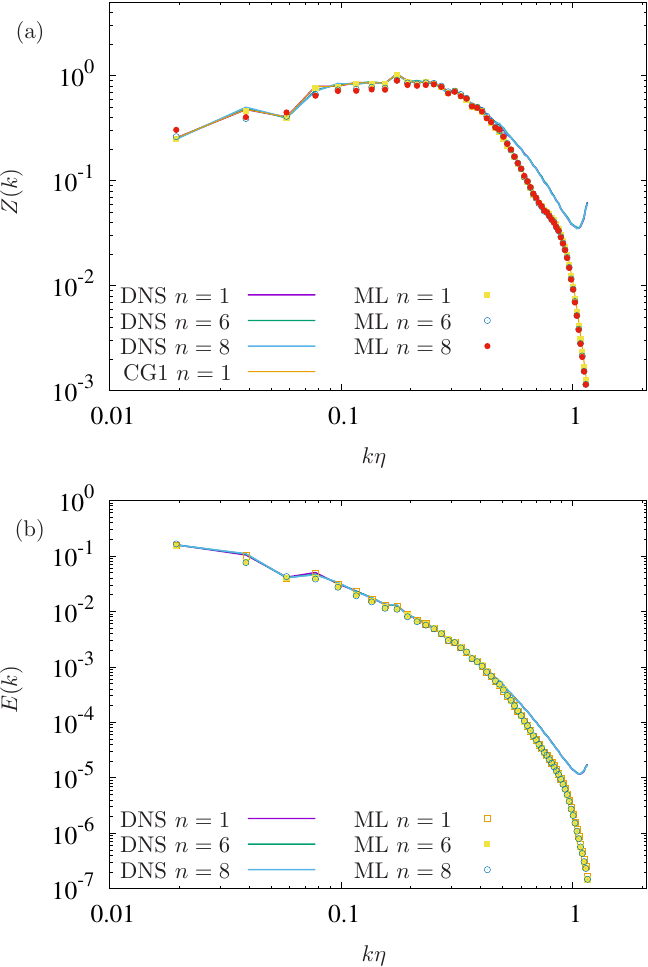}
\end{center}
\caption{(a) Enstrophy spectra $Z(k)$ and (b) energy spectra $E(k)$ at different time instants ($t = t_0+n\tau, n=1, 6, 8$) for the ML data (symbols) and the DNS data (lines).}
\label{ens_spe}
\end{figure}
\begin{figure}
\begin{center}
\includegraphics[width=7.4cm,keepaspectratio]{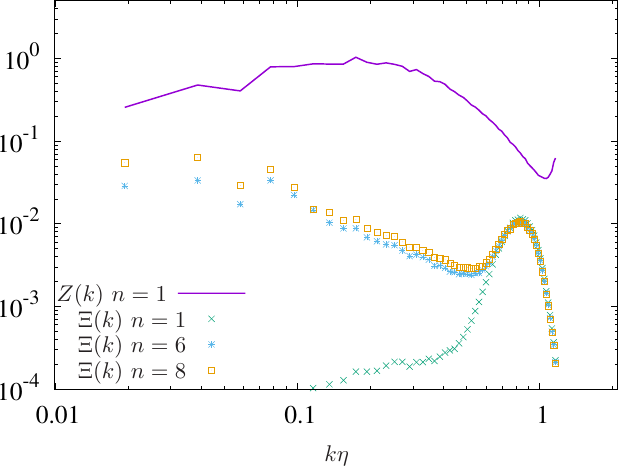}
\end{center}
\caption{The divergence spectra $\Xi(k)$ of the vorticity for the ML data.
The enstrophy spectrum $Z(k)$ for the DNS data at $n=1$ is plotted as a reference.}
\label{div_spe}
\end{figure}
%*************************

Next, we analyze our ML data considering the enstrophy spectrum $Z(k)$, 
which is defined as $Z(k) =(1/2) \sum'_k |{\hat {\bm \omega}}({\bm q})|^2 $ and we have $Z(k)=k^2E(k)$, where $E(k)$ denotes the energy spectrum.
The reason for selecting $Z(k)$ is to catch the statistics in the dissipation range well. 
Here, ${\hat {\bm \omega}}({\bm k})$ is the Fourier transform of ${\bm \omega}({\bm x})$.
We see in Fig. \ref{ens_spe}(a) that the peak of $Z(k)$ where the enstrophy spectra hit the maximum is about $0.2$,
while the external force is imposed in $k\eta \lesssim 0.05$ (see Sec. \ref{sec:dns} and Table \ref{tab1}).
Hence the peak departs from the forcing range.
As $R_\lambda$ increases implying that $\eta$ becomes smaller, a forcing range departs further from the peak (e.g. Ref. \onlinecite{KanedaJoT}).
Figure~\ref{ens_spe}(a) shows that the enstrophy spectra for ML are in almost perfect agreement with those of DNS. 
For wavenumbers $k\eta \gtrsim 0.5$, 
we observe some small discrepancy of $Z(k)$ between ML and DNS. 
We can see that the spectrum for CG1, 
the coarse-grained data at scale $2^{-J+1}$, well overlaps with $Z(k)$ for ML at $n=1$.
This discrepancy can thus be attributed to the wavelet projection where the wavelet coefficients at scale $2^{-J+1}$ are set to zero, as mentioned in Sec. \ref{subsec:ml}.
We therefore conclude that the discrepancy is not crucial.
Figure \ref{ens_spe}(b) plots the energy spectra $E(k)$ for ML and DNS.
We see that the spectra of ML excellently agree with those of DNS.
The observed differences for the enstrophy spectra $Z(k)$ in $k\eta \gtrsim 0.5$ are much reduced for $E(k)$.
Therefore, the enstrophy spectra yield better indicators for the verification of machine learning prediction of vorticity than the energy spectra.

In the LSTM flow prediction using an autoencoder and CNN,
\citet{Nakamura2020} show their predicted energy spectra in the high wavenumber range near by the maximum wavenumber. 
These are enhanced in comparison with the correct spectra for turbulent channel flow, i.e. they overpredict energy. 
This overprediction can be also observed in Refs. \onlinecite{MohanJoT,Shanker2022} for the machine-learning time-evolution prediction using an autoencoder of isotropic turbulence. 

Finally, we examine the influence of the divergence of the vorticity ${\bm \omega}$ for the ML data in which $\nabla {\bm \cdot} {\bm \omega} \ne 0$ in general, 
while for DNS we have $\nabla {\bm \cdot} {\bm \omega}=0$.
The divergent part is given by $ \xi({\bm k}) = \{ {\bm k} {\bm \cdot} {\hat {\bm \omega}}({\bm k}) \} {\bm k}/k^2 $ in ${\bm k}$ space.
Figure~\ref{div_spe} shows the spectra of this part $\Xi(k)$ given as
$\Xi(k) =(1/2) \sum'_k \left| \xi({\bm q}) \right|^2 $.
We can see that the influence of the divergence is not crucial in particular at small scales.
The values of $\Xi(k)$ at large scale, $k\eta \lesssim 0.05$ for later times, are somewhat comparable to the value of $Z(k)$, 
about 25 percent of the magnitude of $Z(k)$. 
However, this issue can be be overcome by using divergence-free biorthogonal wavelets.\cite{DivfreeWavelet, deriaz2007divergence}

%*******************************
\section{Application of WCNN-LSTM to superresolution}\label{SR}
%*******************************
We now apply the pre-trained WCNN-LSTM to regenerate vortices at an instant $t_s (=30.0)$ from a given coarse-grained vorticity field at $t_s$.
This can be regarded as a type of SR applied for recovering the fine-scale flow evolution from coarse-scale predicted flow data.
A key quantity for the degree of the coarse-graining is the normalized wavenumber $k_c\eta$,
where the larger scale data are almost or completely kept for $k \lesssim k_c$ while the smaller scale data are almost or completely lost for $k \gtrsim k_c$. 
One can expect that SR becomes more difficult for decreasing $k_c \eta$.

%++++++++++++++++
\subsection{Coarse-Grained Input Data}
\begin{figure}
\begin{center}
\includegraphics[width=7.5cm,keepaspectratio]{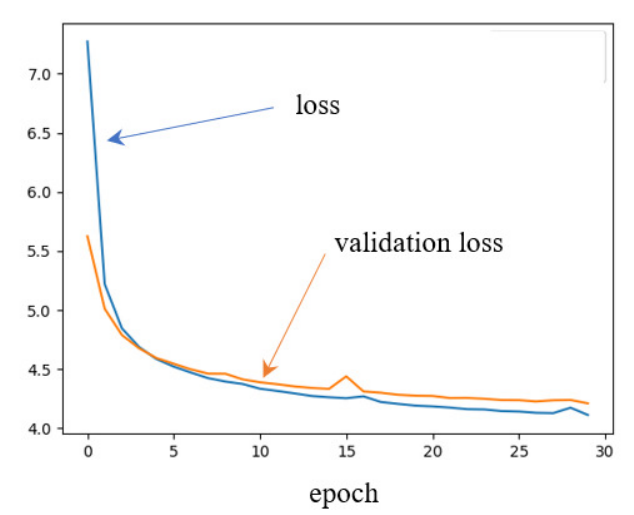}
\end{center}
\caption{Epoch-dependence of loss and validation loss in WCNNSR.} 
\label{errorWCNN}
\end{figure}

%++++++++++++++++
We first obtain a coarse-grained vorticity field from the DNS data.
We apply an FWT, remove the wavelet coefficients at scales $2^{-J+2}$ and $2^{-J+1}$, 
and then reconstruct the flow field from the scaling coefficients at scale $2^{-J+2}$ using an IFWT at scale $2^{-J}$.
We call the obtained coarse-grained vorticity CG2.
We confine ourselves to the same DNS data as those used in our machine learning in Sec. \ref{sec:method},
taking the same periodic boundary conditions imposed on the DNS, as well as the memory limitation into account.
The DNS has been computed at resolution $128^3$ and $k_{\mathrm{max}} \eta \approx 1.16$, as described in Sec. \ref{sec:dns}. 
Thus, we have $J=7$, and the CG2 field lost the data for $k \eta \gtrsim 0.3$.
It is to be noted that the regeneration of vorticity from this type of fields is not learned by our machine learning. 
We then use the SR method developed by Asaka {\it et al}.\cite{asaka2022wavelet} using wavelets, subcube division, and 3D CNN.
We call the method WCNNSR here. 
WCNNSR learns the five snapshots of the DNS data of vorticity at $128^3$ grid points from the training data which are described in Sec. \ref{sec:dns}.
The data at each time instant are divided into 64 subcubes with $32^3$ grid points in physical space.
The subcubes data are then transformed in wavelet space using Coiflet 12 wavelets at one level, by imposing symmetric boundary conditions on the subcubes, the so-called folding technique.
In PyWavelets,\citep{lee2019pywavelets} the boundary condition requires five extra elements of the array in each direction, and thus the size of each subcube becomes finally $8 \times 21^3$ in wavelet space.
Figure \ref{errorWCNN} shows the losses in the WCNNSR as a function of epoch.

We now apply the WCNNSR to the CG2 data at $t=t_s=30.0$,
and then we use the resulting data as an input data.
The CG2 data are much coarse-grained for WCNNSR, 
and therefore WCNNSR does not recover the small-scale flow contributions at least visually well.
It is to be noted that 
the DNS data at $t=t_s$ are not our training data in the WCNNSR and the WCNN-LSTM.
Figure \ref{vis3D_SR2} shows the isosurfaces of $|{\bm \omega}|$ for DNS, the coarse-grained data CG2, the predicted data using WCNN-LSTM including SR (ML), and SR applied to CG2 only (WCNNSR).
The visualization for the DNS data is the same as the DNS at $n=1$ shown in Fig. \ref{vis3D}.
The vorticity tubes observed in the DNS are not well retained in CG2, due to the coarse-graining.
We can see some fragmentation of the tubes in CG2.
In contrast, applying SR to CG2 using  WCNNSR, which is denoted by WCNNSR for simplicity,
we can observe that some tubes regenerated, however much fewer than in DNS.
Moreover, the degree of coarse-graining of the WCNNSR data is almost the same as the CG2 data, as shown in Sec. \ref{ResultSR}.
Therefore, the WCNNSR data are expected to be likewise suitable as input data for WCNN-LSTM. 
Asaka {\it et al}.\cite{asaka2022wavelet} reported that the WCNNSR method works well for less coarse-grained input data at higher $R_\lambda$.
A result of WCNN-LSTM of the CG2 data without WCNNSR is presented in Appendix \ref{appen}.

%-------------------
%-------------------

%+++++++++++++++++
\subsection{Procedures of superresolution using WCNN-LSTM}
%+++++++++++++++++
\begin{figure}
\begin{center}
\includegraphics[width=\linewidth,keepaspectratio]{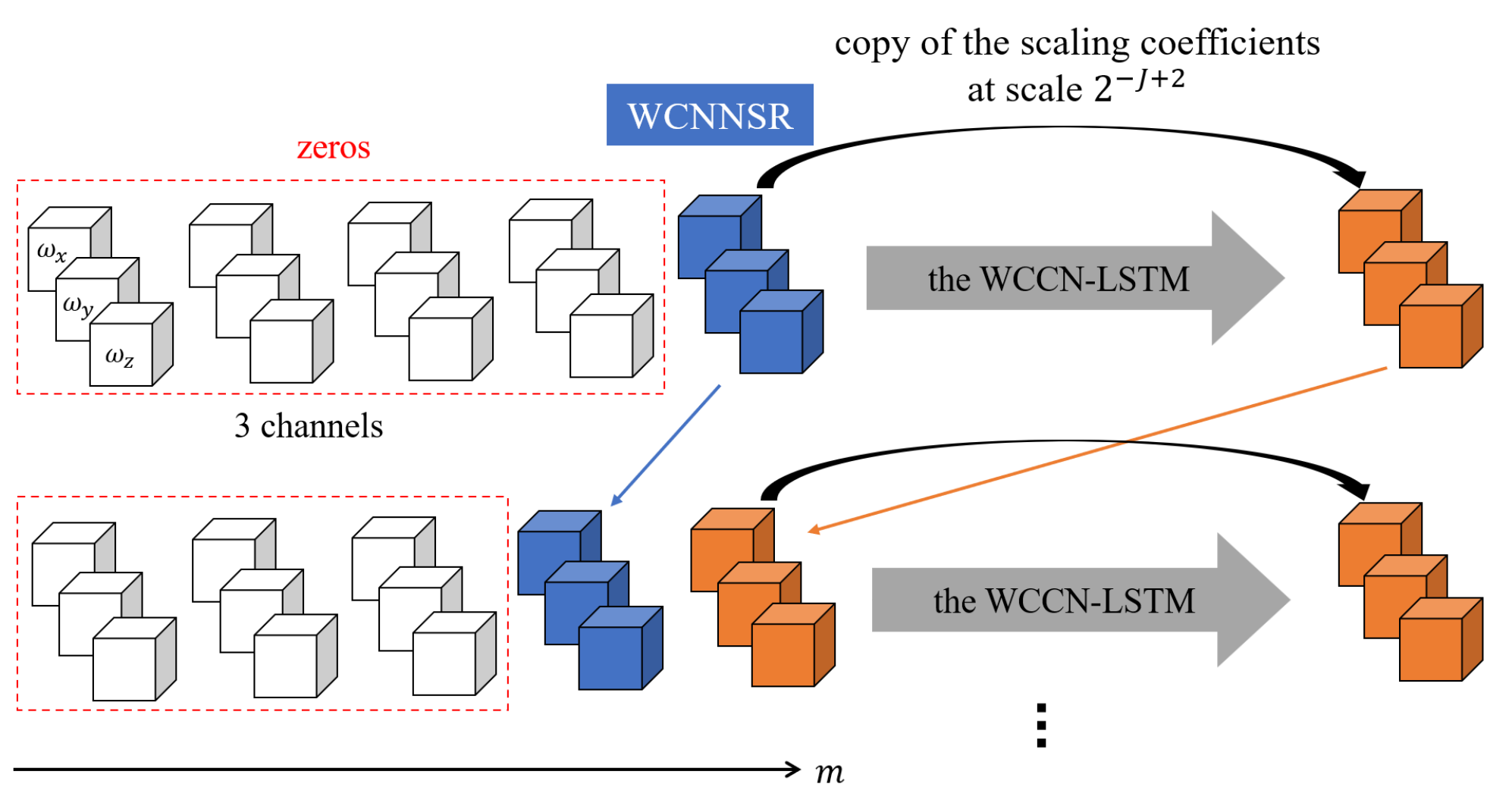}
\end{center}
\caption{SR method based on the pre-trained WCNN-LSTM.} 
\label{WCNN-LSTMSRmethod}
\end{figure}
%++++++++++++++++++
Figure~\ref{WCNN-LSTMSRmethod} illustrates our SR procedure using the pre-trained WCNN-LSTM, introduced in Sec. \ref{subsec:ml}. 
Our input data are four empty data sets whose values are zero for pseudo-time instants $t_s + m \Delta \tau$ ($m=-4,-3,-2,-1$) together with the predicted vorticity data obtained by WCNNSR at time $t_s$,
where $ \Delta \tau \approx 0.5 \tau_{\eta}$.
We have chosen the four empty data sets such that the influence of the input data at the instants given by negative indices $m$ on the SR at $t_s$ can be reduced. 
WCNN-LSTM predicts the scaling coefficients of vorticity at scale $2^{-J+1}$ and at time $t_s + \Delta \tau$.
Using the FWT with periodic boundary conditions, we decompose them into the scaling coefficients at scale $2^{-J+2}$ and the wavelet coefficients at scale $2^{-J+2}$. 
Then, we copy the scaling coefficients at scale $2^{-J+2}$ of the input WCNNSR data at $t_s$ to the scaling coefficients at scale $2^{-J+2}$ which have been obtained in the above decomposition.
Therefore, the information corresponding to the wavelet coefficients at scale $2^{-J+2}$ evolves in our SR results.
Afterwards using the IFWT, 
we obtain the output data in physical space at $t_s+\Delta \tau$. 
They are then used as input data in the next step,
in addition to three empty data sets at $t_s + m \Delta \tau$ ($m=-3,-2,-1$) and one WCNNSR data at $t_s$.
We obtain the output data at $t_s+2 \Delta \tau$.
The input data in the $\ell$-th step are the data $t_s + m \Delta \tau$ ($m=-4+\ell,-3+\ell,-2+\ell,-1+\ell$),
where $\ell \in \mathbb{N}$.
The data at negative indices $m$ are empty, the data at $m=0$ are the WCNNSR data, and the data at positive indices $m$ are predicted by the WCNN-LSTM with the above-mentioned copy of the scaling coefficients at scale $2^{-J+2}$ of the WCNNSR data.
The data at $t_s + \ell \Delta \tau$ are obtained by the same procedure described above.
Eventually we get the output data in physical space using IFWT with the periodic boundary conditions at each pseudo-time instant.

%+++++++++++++++++
\subsection{Results of the superresolution} \label{ResultSR}
%+++++++++++++++++
\begin{figure}
\begin{center}
\includegraphics[width=12cm,keepaspectratio]{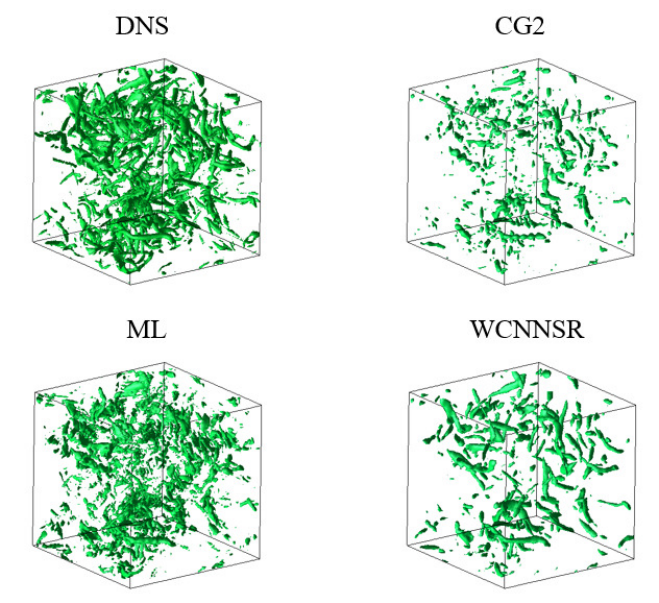}
\end{center}
\caption{Visualization of isosurfaces of $|{\bm \omega}|$ at $|{\bm \omega}|=\omega_m + 2.5 \omega_\sigma$ for the DNS data at $t=t_s$, the CG2 data at $t=t_s$, the SR data obtained by WCNN-LSTM, which is denoted by ML, at $m=4$,
and the WCNNSR data at $t=t_s$. 
Here, $\omega_m$ and $\omega_\sigma$ are the mean value and the standard deviation of $|{\bm \omega}|$ for the DNS data, respectively. 
The grid resolution is $128^3$. }
\label{vis3D_SR2}
\end{figure}
%****
\begin{figure}
\begin{center}
\includegraphics[width=12cm,keepaspectratio]{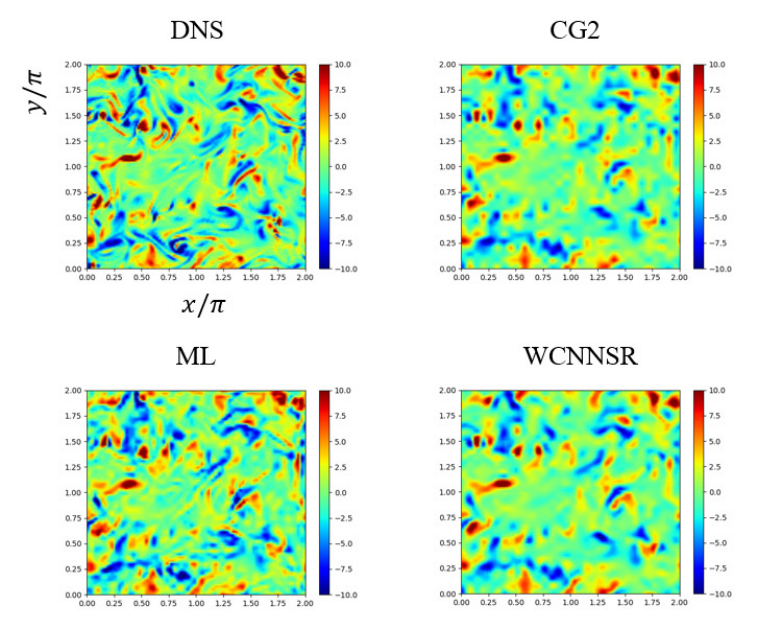}
\end{center}
\caption{Visualization of the vorticity components $\omega_x$ on an $x-y$ plane for the DNS data at $t=t_s$, the CG2 data at $t=t_s$,
the ML data at $m=4$,
and the WCNNSR data at $t=t_s$. }
\label{vis2D_SR}
\end{figure}
%******************

Now, we verify the SR using the WCNN-LSTM to recover fine-scale information of the flow.
Figure \ref{vis3D_SR2} shows the isosurfaces of vorticity magnitude $|{\bm \omega}|$.
ML denotes the visualization for the SR of the WCNNSR result, by using the WCNN-LSTM.
The ML results seem to well preserve most of the positions of the tubes in DNS,
though we can observe some differences between the visualizations of DNS and ML:
the isosurfaces of DNS are smoother than those of ML.
Figure \ref{vis2D_SR} visualizes $\omega_x$ on an $x-y$ plane.
We can see that ML reproduces the vorticity distribution of DNS well,
though there are again some differences between ML and DNS looking at details.
Table~\ref{ensSR} shows the enstrophy ${\mathcal{Z}}$ for DNS, CG2, WCNNSR, and ML.
The enstrophy ${\mathcal{Z}}$ for ML is comparable to ${\mathcal{Z}}$ for DNS,
while the other values for CG2 and WCNNSR data are much reduced. 

\begin{figure}
\begin{center}
\includegraphics[width=7.5cm,keepaspectratio]{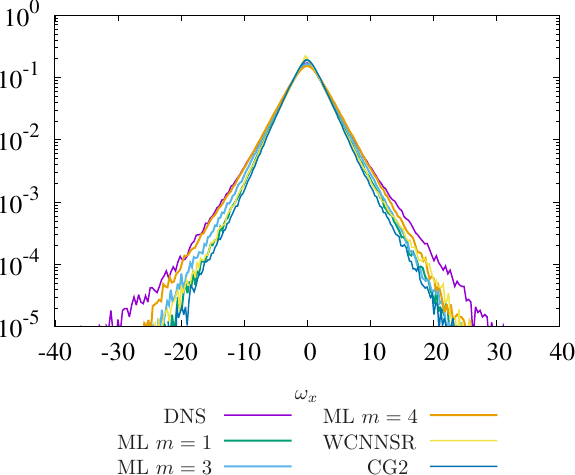}
\caption{PDFs of $\omega_x$ for DNS, ML ($m=1, 3, 4$), WCNNSR and CG2.}
\label{SRox}
\end{center}
\end{figure}

\begin{figure}
\begin{center}
\includegraphics[width=6cm,keepaspectratio]{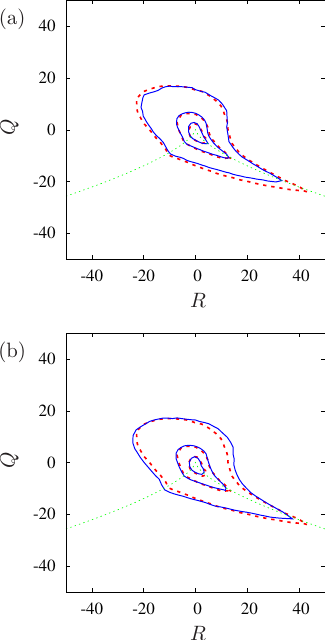}
\caption{Joint PDFs of $Q$ and $R$ for (a) DNS and WCNNSR, and for (b) DNS and ML $(m=4)$.
The contour lines for DNS, WCNNSR, and ML are denoted by the red dashed lines, black solid lines, and the blue solid lines, respectively.
The contour lines for all cases are set to $10^{-4}$, $10^{-3}$ and $5 \times 10^{-3}$, starting near the origin. 
The green dotted lines represent $27R^2/4 +Q^3=0$.
}
\label{SRqr}
\end{center}
\end{figure}

We move on to the statistics of small-scale quantities.
Figure \ref{SRox} gives a plot of PDFs of $\omega_x$ for the different fields.
We observe that the PDFs for WCNNSR and CG2 are much narrower than the PDF for DNS.
We can also see that the PDF of ML becomes wider as the pseudo-time progresses.
We find that the PDF of ML at $m=4$ (orange line) is in good agreement with that of DNS,
though the former is somewhat narrower than the latter.
In Fig.~\ref{SRqr}, we compare the joint PDFs of the second and third invariant of the velocity gradient tensor, $Q$ and $R$, for WCNNSR and ML ($m=4$) with those of DNS.
We can see that the PDFs of WCNNSR and ML well agree with the joint PDF of DNS.
The departure from the DNS results can be seen only at the smallest isoline value, $10^{-4}$.

\begin{figure}
\begin{center}
\includegraphics[width=7.5cm,keepaspectratio]{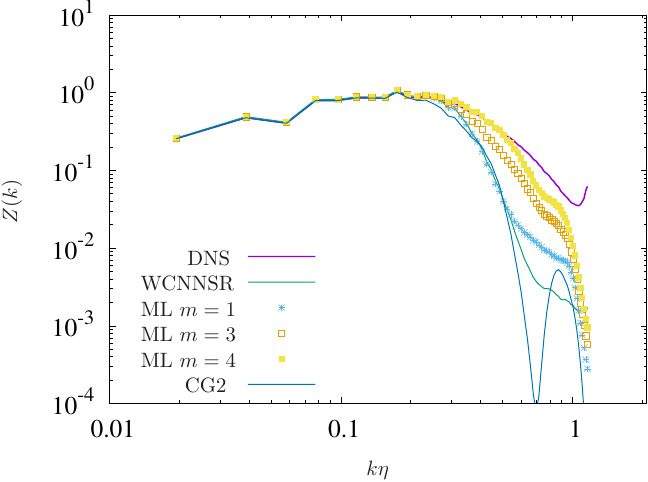}
\end{center}
\caption{Enstrophy spectra $Z(k)$ vs $k \eta$ for DNS, ML ($m=1, 3, 4$), WCNNSR and CG2.}
\label{SR_Zspe}
\end{figure}

\begin{figure}
\begin{center}
\includegraphics[width=7.4cm,keepaspectratio]{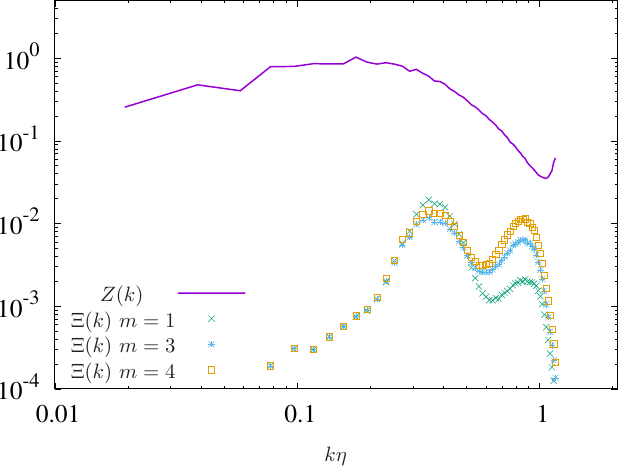}
\end{center}
\caption{The divergence spectra $\Xi(k)$ vs $k \eta$ for ML ($m=1, 3, 4$). 
The enstrophy spectrum $Z(k)$ of the DNS data at $t_s$ is plotted as a reference.}
\label{SR_divpart}
\end{figure}

To get deeper insight into the small scales,
we examine the enstrophy spectra $Z(k)$.
In Fig.~\ref{SR_Zspe}, we can see that 
for CG2 and WCNNSR, the enstrophy spectra are much reduced for $k\eta \gtrsim 0.2$.
We find that the degree of coarse-graining of the WCNNSR data is almost the same as that of the CG2 data in terms of $Z(k)$.
The values of $Z(k)$ for ML for $k\eta \gtrsim 0.2$ grow in the pseudo-time,
and almost saturate at $m=3$ and $m=4$.
We can see that $Z(k)$ for ML at $m=4$ excellently agrees with that for DNS.
We recall that 
we copy the scaling coefficients at scale $2^{-J+2}$ ($J=7$) of the WCNN data to the output data every pseudo-time instant in the SR method using the WCNN-LSTM prediction.
Therefore, the enstrophy spectra of ML are almost the same as $Z(k)$ for WCNNSR for $k\eta \lesssim 0.2$. 
Figure~\ref{SR_divpart} shows the divergence spectra of the vorticity for ML. 
This confirms that the influence of divergence is not crucial.

%******************
\begin{table}
\caption{Enstrophy of the DNS data at $t=t_s$, the CG2 data at $t=t_s$,
the ML data at $m=4$, and the WCNNSR data at $t=t_s$.}
\label{ens_tab}
\begin{center}
\begin{tabular}{*{5}{c}}
\hline
 & DNS & CG2 & WCNNSR & ML  \\
\hline
${\mathcal{Z}}$ & $20.0$ & $13.1$ & $14.4$ & $19.4$ \\
\hline
\end{tabular}
\label{ensSR}
\end{center}
\end{table}
%******************
%---
%******************
%******************

%-----------------------------
\section {Conclusions}\label{sec:conclusions}
%+++++++++++++++++++++++++++++
We have developed a wavelet-based machine learning method (WCNN-LSTM) for predicting the time evolution of vorticity for homogeneous isotropic turbulence where vortex tubes are preserved.
To this end, we combined two neural network architectures, 3D CNN and LSTM. 
The latter permits learning the time evolution, 
while the former trains the network to reproduce not only the turbulent flow but also the pronounced and well-localized vortex tubes.
We have used compactly supported orthogonal wavelets. 
By construction wavelets well catch the information of position and scale of the fields.
The wavelets efficiently represent multi-scale and intermittent fields, here turbulent vorticity fields exhibiting vortex tubes.\cite{Farge2001,okamoto2007coherent}
The fast wavelet transform and its inverse allow switching rapidly between physical space and wavelet coefficient space.
Wavelet projection is then used to reduce the size of the input data.
Therefore, the projection reduces the memory required for learning.
We applied the developed WCNN-LSTM to DNS data of homogeneous isotropic turbulence in a $(2\pi)^3$ periodic box at the Taylor microscale Reynolds number $92$ computed at resolution $128^3$ and $k_{\mathrm{max}} \eta \approx 1.16$. 
The dimensionless wavenumber $k \eta$ where the enstrophy spectra hit the maximum is about $0.2$,
while the external force is imposed in $k\eta \lesssim 0.05$. 
The results predicted by WCNN-LSTM have been assessed in comparison with the DNS data.
The predicted fields satisfy periodic boundary conditions by using the IFWT.
Visualization of isosurfaces of vorticity magnitude showed that 
vortex tubes are well retained in the WCNN-LSTM flow prediction.
The PDFs of vorticity and velocity, which are predicted by WCNN-LSTM, well agree with those of the DNS data at time instants larger than $3\tau_\eta$,
where $\tau_\eta$ is the Kolmogorov time-scale,
The flow topology characterized by second and third invariants of the velocity gradients is likewise well retained by WCNN-LSTM.  
We employed the enstrophy spectra to get deeper information into the dissipation range. 
We observed that the predicted enstrophy spectra well agree with those for the DNS.
The influence of the divergence of the predicted vorticity was shown to be negligible.
As expected
we found that the enstrophy spectra are more suitable than the energy spectra for verification of machine learning prediction of vorticity.

Then, we applied our pre-trained WCNN-LSTM model to the SR of a coarse-grained vorticity of the DNS data at a time instant $t_s$.
The information at $ k\eta \gtrsim 0.2$ is much reduced in the coarse-grained data, 
and the vortex tubes of the DNS data at $t=t_s$ are lost in the coarse-grained vorticity.
We showed that the vortex tubes are well regenerated from the coarse-grained vorticity by the SR,
though the previously developed wavelet-based SR model at a time instant\cite{asaka2022wavelet} failed in the SR of the coarse-grained vorticity. 
It is to be noted that the WCNN-LSTM model was not trained to learn the regeneration.
The vortex tubes are visualized by using isosurfaces of the modulus of the vorticity.
We can see that the isosurfaces for the predicted vorticity well agree with those of the DNS data, 
though the latter are smoother than the former.
Further improvement of the quality of SR using machine learning of the time evolution of vorticity remains an interesting issue for future studies.

Predicting the dynamics in fully developed turbulent flow with DNS is a costly endeavor, 
because the number of degrees of freedom increases approximately with $R_\lambda^{9/2}$. 
Machine learning of turbulent flows thus requires much more memory and computational cost, as the Reynolds number increases.
The memory and cost could be reduced by multi-level wavelet decomposition with or without nonlinear wavelet filtering. 
%before: wavelet nonlinear filtering
In machine learning based on multi-level wavelet decomposition, 
the wavelet coefficients at each scale are learned, 
which means scale-by-scale machine learning.
In Farge {\it et al}.,\cite{Farge1999} coherent vortex simulation was proposed to compute the time evolution of the coherent vorticity, 
while neglecting the influence of the incoherent flow to model turbulence dissipation. 
The coherent vorticity consists of few intense wavelet coefficients of vorticity, 
and the intense coefficients are extracted by using wavelet nonlinear filtering.\cite{Farge2001,okamoto2007coherent}
Tracking the time evolution of the coherent vorticity in wavelet space while reducing the required memory needs adaptive computation based on wavelets.
The application of the adaptive computation to the machine learning remains an open issue.
Moreover, the concept of time parallelization (e.g., Ref. \onlinecite{TimeParallel}) could be useful for efficient learning longer time evolution of turbulence.
We look forward to parallelization of machine learning that will resolve the present limitation of the GPU memory and the computational time.
It could be also interesting to examine whether machine learning can reconstruct the time evolution of 3D turbulent flows from 2D slices of the flow. 
The prediction of time evolution of turbulence with different dynamics, e.g., inertial particles can be interesting on the basis of Oujia {\it et al}. \cite{oujia2022ctr,maureloujia2023neural} for synthesizing preferential concentration of particles in isotropic turbulence.

\smallskip

\begin{acknowledgments}
This work used computational resources of the supercomputer ``Flow" provided by Information Technology Center, Nagoya University, partially through the HPCI System Research Project (Project ID: hp230143).
KS acknowledges partial support by the French Federation for Magnetic Fusion Studies (FR-FCM) and the Eurofusion consortium, funded by the Euratom research and training programme 2021-2022 under grant agreement No 633053. The views and opinions expressed herein do not necessarily reflect those of the European Commission.
We also acknowledge T. Maurel-Oujia for fruitful discussion on this paper.
\end{acknowledgments}

\appendix
\section{WCNN-LSTM without WCNNSR}\label{appen}
\begin{figure}
\begin{center}
\includegraphics[width=7.5cm,keepaspectratio]{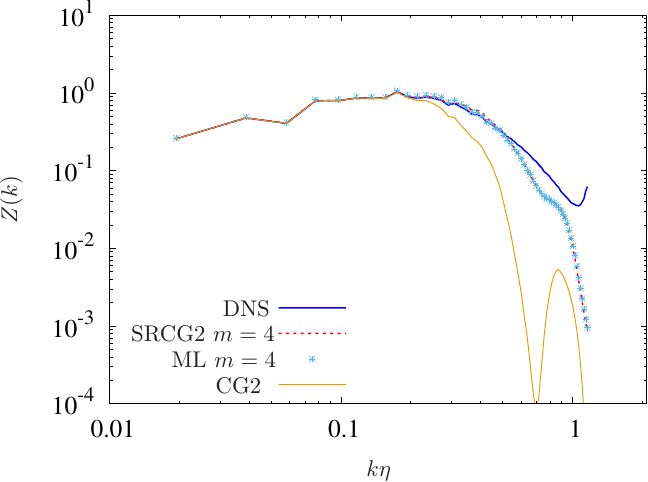}
\end{center}
\caption{Enstrophy spectrum $Z(k)$ vs $k \eta$ for SRCG2 at $m=4$ together with the enstrophy spectra for DNS, ML at $m=4$, and CG2.}
\label{SR_appe}
\end{figure}

We shortly describe the SR of the CG2 data by using the WCNN-LSTM without WCNNSR.
This SR is here denoted by SRCG2.
We select the enstrophy spectra $Z(k)$ here.
The spectra provide good verification of the results in the dissipation range, as discussed in Sec. \ref{ResultSR}. 
Figure \ref{SR_appe} shows that 
$Z(k)$ for SRCG2 at $m=4$ takes somewhat larger values than the $Z(k)$ of DNS at $m=4$ for $k \eta \approx 0.4$.

\providecommand{\noopsort}[1]{}\providecommand{\singleletter}[1]{#1}%
%

%\bibliography{AYS.bib}

\end{document}